# Projected Performance Advantage of
# Multilayer Graphene Nanoribbon as Transistor Channel Material

Yijian Ouyang[1], Hongjie Dai[2], and Jing Guo[1]

[1] Department of Electrical and Computer Engineering, University of Florida, Gainesville, FL, 32611

[2] Department of Chemistry, Stanford University, Stanford, CA, 94305

## ABSTRACT

The performance limits of the multilayer graphene nanoribbon (GNR) field-effect transistor (FET) are assessed and compared to those of monolayer GNR FET and carbon nanotube (CNT) FET. The results show that with a thin high-$\kappa$ gate insulator and reduced interlayer coupling, multilayer GNR FET can significantly outperform its CNT counterpart with a similar gate and bandgap in terms of the ballistic on-current. In the presence of optical phonon scattering, which has a short mean free path in the graphene-derived nanostructures, the advantage of the multilayer GNRFET is even more significant. The simulation results indicate multilayer GNRs with incommensurate non-AB stacking and weak interlayer coupling are the best candidate for high performance GNR FETs.

# I. Introduction

Graphene [1-3] is a monolayer of carbon atoms packed into a honeycomb lattice, which possesses atomically thin body and an area scale orders of magnitude greater, making it an ideal two-dimensional (2D) system. A 2D graphene is a semimetal without a bandgap, but a bandgap opens if a field-effect transistor (FET) channel is built on a nanometer-wide graphene nanoribbon (GNR) due to the width direction confinement [4], which leads to subband formation as well [5]. The high mobility (up to 200,000 $cm^2/Vs$) and carrier velocity (~$10^8$ cm/s) demonstrated in 2D graphene have stimulated strong interests on graphene electronics [1]. The transport properties of GNRs in experiments have been so far hindered by imperfect edges [6,7], but excellent transport properties have been theoretically predicted for structurally perfect GNRs [8]. The GNRFET, however, suffers from the problem of a low on-current due to its nanometer-wide channel, and it has been shown previously that the ballistic performance limits of a monolayer GNRFET are not better than a CNTFET in terms of the on-current and on-off current ratio [9].

In this study we assess the performance limits of GNRFETs and CNTFETs using a well established ballistic transistor model, which has been applied to various two-dimensional and one-dimensional channel transistors before [10]. The performance limits of a transistor are achieved when the contacts are ideal and the channel is ballistic (no scattering). Schottky barriers can play an important role in CNTFETs [11] and GNRFETs. The Schottky barrier is known to lower the on-current and increase the off-current due to ambipolar *I-V* characteristics. These detrimental effects in the Schottky barrier CNTFETs, however, can be eliminated by using a metal-oxide-semiconductor (MOS) FET device structure [12], which has heavily doped source and drain extensions as shown in Fig. 1(a). The performance limits are therefore assessed for a ballistic MOSFET structure with semi-infinite source and drain extensions. In this condition, the simple semiclassical model agrees with detailed quantum mechanical transistor simulations for a channel length down to about 10 nm [10], as long as the direct source-drain tunneling current is relatively small compared to the total source-drain current.

Multilayer GNR were demonstrated in recent experiments [13,14]. We show that the modeled ballistic performance limits of a multilayer GNRFET can be significantly better than its corresponding CNTFET in terms of the on-current and on-off current ratio. The advantage becomes even bigger in the presence of optical phonon scattering, which is known to be strong in graphene-derived nanostructures. The important role of gating technology and interlayer coupling are investigated for achieving the performance advantage of multilayer GNRFETs over CNTFETs. We found that weak interlayer coupling, which is accessible through incommensurate non-AB stacking, is desired for high performance multilayer GNRFET.

## II. Approach

The modeled graphene nanoribbon and carbon nanotube have similar bandgaps for a fair comparison. The (20, 0) zigzag CNT is semiconducting, which results in a diameter of $d_{CNT} \approx 1.6$ nm and a bandgap of $E_g \approx 0.50$ eV. The $n=22$ armchair edge GNR (where $n$ denotes the number of carbon dimmer lines [15]) is also semiconducting, which results in a width of $W_{GNR} \approx 2.7$ nm and a bandgap of $E_g \approx 0.52$ eV for a monolayer GNR. Both transistors have doped source and drain extensions, as schematically shown in Fig. 1(a). For the GNRFET, a wrapped-around gate is used as shown in Fig. 1(b). For the CNTFET, a coaxial gate is used as shown in Fig. 1(c). The nominal gate insulator is a 3-nm-thick $ZrO_2$, which has a relative dielectric constant of $\kappa \approx 25$. A power supply voltage of $V_{DD}$=0.5 V is used.

The bandstructures of the CNT and GNR channels are required as an input to the transistor model. A nearest-neighbor $p_z$ tight binding approach was used. The hopping integral is taken as $t_{cc}$=-2.7 eV. The edge effect plays an important role in the GNR, and a factor of 1.12 is used for the hopping parameter between two edge carbon atoms to count the edge relaxation effect [16]. For the AB-stacking structure of the multilayer graphene as shown in Fig. 2, an interlayer coupling of 0.3 eV is set only for vertically aligned two atoms in two neighboring layers [17]. The nominal device has an interlayer coupling of zero, which is the lower limit of interlayer coupling. Weak interlayer coupling could be achieved by incommensurate non-AB-stacking structures, and it is most preferable for better device performance as discussed below. The nominal value is also varied to examine the effect of interlayer coupling strength on the device performance.

A "top-of-barrier" transistor model is used to assess the performance limits of the multilayer GNRFETs. The model fills the $+k$ states at the top of the barrier with the source Fermi level $E_{FS}$ and the $-k$ states with the drain Fermi level $E_{FD}$ as shown in Fig. 3(a). Self-consistent electrostatics is treated by a simple capacitance model, in which the gate insulator capacitance value is computed by a numerical Poisson solver and source- (drain-) channel capacitance is ignored for simplicity. The optical phonon scattering in carbon nanotubes [18] or graphene nanoribbons has a short mean free path of ~10 nm, which could play an important role even for transistors with a sub-100nm channel length. We also consider the transistor performance in the presence of the OP scattering by setting an effective Fermi level at $E_F^{'}=E_{FS}-\hbar\omega_{OP}$, which determines the population of the -$k$ states in the presence of OP scattering, as shown in Fig. 3(b).

$\hbar\omega_{OP}\approx0.18$ eV is the optical phonon energy in CNTs and GNRs.

## III. Results

Because the transistor performance strongly depends on the bandstructure of the channel material, we first examine the bandstructures of the multilayer GNRs. The upper left panel of Fig. 4(a) plots the bandstructure of the monolayer $n=22$ AGNR, which has a width of $W_{GNR}\approx2.7$ nm and a band gap of $E_g\approx516$ meV. Compared to a CNT in which a periodic boundary condition applies in the circumferential direction of the CNT and a valley degeneracy factor of 2 exists for each band, the monolayer AGNR does not have valley degeneracy due to a different quantization boundary condition in the width direction of the GNR. At the limit of zero interlayer coupling, an $m$-layer GNR has the same bandstructure as the monolayer, but each bands becomes a factor of m degenerated due to $m$ uncoupled layers. The interlayer coupling lifts the subband degeneracy. Figure 4(a) also shows the bandstructure s of 2-, 5- and 10- layer multilayer GNRs with an AB stacking structure. As the number of the layer increases, the subband spacing reduces, and the upper subbands should become more accessible for carrier transport. Furthermore, as the layer number increases from 1 to 10, the bandgap decreases from 516 meV to 197 meV. We also examined the dependence of the bandgap on the interlayer coupling strength for the multilayer GNRs as shown in Fig. 4(b). As the interlayer coupling strength increases from the zero limit to the value for the AB stacking structure, the bandgap monotonically decreases regardless of the number of the GNR layers. The decrease is more significant for a multilayer GNR with a larger number of layers.

Next, we examine the effect of various gating on transistor performance by varying the SiO$_2$ gate oxide thickness $t_{ox}$ from 0.23 nm to 20 nm, as shown in Fig. (5). A common off-current of 10 nA is specified for all transistors. The lower limit of the simulated SiO$_2$ thickness is an equivalent case of a 3-nm-thick ZrO$_2$ gate for the modeled gating structure as shown in Fig. 1, which results in the same gate insulator capacitance. The on-current increases only slightly by 12% as the number of the layers increases from 1 to 10 at $t_{ox}$ =20 nm as shown in Fig. 5(a). The transistor on-current is determined by the product of the charge density and the average carrier velocity at the top of the channel potential barrier. The relatively small improvement is largely due to a small increase in carrier density as the number of layers increases. The gate to channel capacitance $C_g$ is a serial combination of insulator capacitance $C_{ins}$ and the quantum capacitance $C_q$, and it is limited by $C_{ins}$ since it is much smaller than $C_q$. A larger number of layers results in a slightly larger $C_{ins}$ and thereby slightly larger $C_g$ due to a thicker channel. For thick oxide, the advantage of using a multilayer GNR channel in terms of the on-current is insignificant.

In contrast, the advantage of the multilayer GNR channel in terms of the on-current becomes more significant as $t_{ox}$ decreases, as shown in Fig. 5(a). Table 1 compares the performance of various channels with a high-κ 3 nm ZrO$_2$. For a common off-current of 10 nA, the monolayer GNR channel has a slightly larger gate insulator capacitance, however a 13% smaller on current than the CNT channel, because of the lack of valley degeneracy in the monolayer GNR. As the number of the layer increases, the on-current surpasses that of the CNT. For a 5-layer GNR, the on-current is 97% larger than that of the CNT, and for a 10-layer GNR, the on-current is 180% larger than that of the CNT. For a similar gating technology, the ballistic on-current of the multilayer GNR can be significantly larger than the CNT with a similar bandgap.

In order to understand the factors that contribute to the increase of the on-current, we also computed the gate insulator capacitance and the average carrier velocity as shown in Table 1. The average carrier velocity decreases from $4.28\times10^7$ cm/s to $3.50\times10^7$ cm/s as the number of GNR layers $m$ increases from 1 to 10 because more carriers populate closer to the bottom of the $m$-fold degenerate lowest subbands where the band-structure-limited velocity is low. The on-current, however, increases by a factor of 3.2 due to the increase of the gate capacitance $C_g$ by a factor of 3.9. It is also interesting to notice that the increase of the gate capacitance significantly outpaces the increase of the gate insulator capacitance stemming from a thicker GNR body as the number of the GNR layer increases, because of the proportional increase of the quantum capacitance as a function of the number of the layers. Furthermore, for the modeled thin high-κ gate insulator, the quantum capacitance and the gate insulator capacitance becomes comparable, and both of them play a role in determining the gate capacitance.

The improvement of on-current is even more significant in the presence of OP scattering. As shown in Table 2, the on-current increases even more significantly as the number of layers increases in the presence of OP scattering. The on-current of the 10-layer GNR channel is 260% larger than that of the CNT. In addition, the average carrier velocity is also larger, which is different from the case of the ballistic channel as shown in Table 1. For a CNTFET, the OP scattering results in a saturation current close to $(4e^2/h)\times(\hbar\omega_{OP}) \approx 28\mu A$, where the OP energy $\hbar\omega_{OP} \approx 180$ meV when only the 1st subband conducts the current, which is the case for the CNTFET because $E_{FS} - E_1 \approx 216$ meV is smaller than the spacing between the 1st and the 2nd subbands $E_2 - E_1 \approx 226$ meV. The OP scattering results in considerably increased population of the $-k$ states, which also lowers the average carrier velocity at the top of the barrier from $4.61\times10^7$ cm/s to $2.72\times10^7$ cm/s. In contrast, the effect of OP scattering on the 10-layer

GNRFET is small, because the value of $E_{FS} - E_1 \approx 120$ meV is smaller than the OP energy. In a multilayer GNR, more subbands are responsible for delivering the on-current, and therefore, the $E_{FS}$-$E_1$ value is smaller. As a result, the OP scattering only has a small effect on the on-current and average carrier velocity of the 10-layer GNRFET. Different from the case of the ballistic channel, the average carrier velocity of the 10-layer GNRFET is also about 20% larger than that of the CNTFET in the presence of OP scattering, which promises faster intrinsic transistor speed.

Finally, we examine the dependence of the on-current on the interlayer coupling. As shown in Figure 7, the largest on-current is achieved at the zero interlayer coupling and the on-current decreases as the interlayer coupling increases. The ballistic on-current monotonically decreases from 51 µA to 36 µA for 2 layers and from 77 µA to 58 µA for 5 layers as the interlayer coupling increases to the value of AB stacking. The reason is that increase of the interlayer coupling increases the spacing between the subbands, which makes the higher subbands more difficult to be accessed for delivering the on-current. In addition, the increase of interlayer coupling also decreases the bandgap, as shown in Fig. 4. As the bandgap decreases, the band-to-band tunneling can be turned on, which can significantly increase the off-current. (Modeling the band-to-band tunneling current is beyond the capability of the semiclassical model used here.) Non-AB-stacking multilayer structures have been recently observed in experiments for both GNRs unzipped from CNTs [13] and CVD-grown multilayer graphene structures. Weakening the interlayer coupling by non-AB-stacking structures, such as the randomly stacking structures, should be pursued for boosting the performance of the multilayer GNRFETs.

### IV.    Discussions and Conclusions

The performance of the multilayer GNR MOSFETs can also be compared to that of the silicon MOSFETs. Such comparison is clouded by the different dimensionality of the channel material, and we simply discuss how dense a GNR array channel needs to be to meet the performance goal at the end of the ITRS roadmap [19]. The ITRS roadmap calls for an on-current of about 2700 µA/µm and an off-current of 0.60 µA/µm at a power supply voltage of 0.65 V for the technology nodes near year 2020. To reach the on-current of 2700 µA/µm, a 10-layer array GNRFET with 26 GNRs per µm is needed for the on-current as shown in Table 2. The array has an off-current of 0.26 µA/µm, which is less than a half of the ITRS goal, although the power supply used in the GNRFET simulation is only 0.5 V. A recent experimental study characterized the mobility degradation of monolayer graphene on a $SiO_2$ substrate [20]. Another experimental study showed that the multilayer graphene transistor is more immune to the noise compared to the monolayer graphene transistor [21], whose one-atom-thick body is susceptible to the charge impurities and oxide traps near the transistor channel. Therefore the multilayer GNRFET is expected to have a better immunity to the adverse effects from substrates than the monolayer GNRFET.

The bandstructure of a multilayer GNR is assumed to be gate-voltage-independent in this study. Large electric field and a significant potential drop between the layers, however, can alter the bandstructure of a multilayer GNR. For the simulated gate-all-around structure and the low applied voltage below 0.5V, the electric field between the layers and its resulting potential drop between the layers is small (which is less than 30 meV as computed by a separate self-consistent atomistic simulation as described in Ref.[22,23]). Its effect on the transistor *I-V* characteristics is negligible for the modeled gating structure and power supply voltage.

In summary, we have shown the important role of developing good gating technology and weakening the interlayer coupling for improving the performance of multilayer GNRFETs. The thin high-$\kappa$ gate insulator is already in production for Intel 45nm transistor, and its application to GNRFETs remains to be developed. Weakening the interlayer coupling could be experimentally achieved in non-AB-stacking multilayer graphene. The performance limit of a well-designed multilayer GNRFET can significantly outperform its CNT counterpart with a similar gate and bandgap in terms of the on-current, at either the ballistic limit or in the presence of OP scattering.

**Acknowledgement**

This work was supported by the National Science Foundation (NSF) and the Office of Naval Research (ONR), Intel, and MARCO MSD.

**Tables**

**Table 1** Comparison of the (20, 0) CNTFET to 1-, 5-, and 10-layer *n*=22 GNRFETs at the ballistic performance limits. The transistor structures are shown in Fig. 1. The gate insulator thickness is 3 nm and the dielectric constant is 25 (for $ZrO_2$). The off current ($I_{OFF}$), gate insulator capacitance ($C_{ins}$), on-current ($I_{ON}$), the spacing between the source Fermi level and the top of the 1$^{st}$ subband barrier ($E_{FS}$-$E_1$), the on-off current ratio ($I_{ON}/I_{OFF}$), and the average carrier velocity at the top of the barrier ($\langle v(0) \rangle$) are compared. For the modeled 1-layer GNR, the spacing between the 1st and 2nd subband is $E_2$- $E_1$= 0.121 eV. For the modeled CNT, $E_2$- $E_1$= 0.226 eV.

|  | $I_{OFF}$ (nA) | $C_{ins}$ (F/m) | $I_{ON}$ (uA) | $E_{FS}$-$E_1$ (eV) | $I_{ON}/I_{OFF}$ | $\langle v(0) \rangle$ cm/s |
|---|---|---|---|---|---|---|
| CNT | 10 | 8.42e-10 | 39.9 | 0.233 | 3900 | 4.28e7 |
| 1-layer GNR | 10 | 8.98e-10 | 34.1 | 0.279 | 3400 | 4.61e7 |
| 5-layer GNR | 10 | 1.27e-9 | 77.3 | 0.155 | 7700 | 3.69e7 |
| 10-layer GNR | 10 | 1.73e-9 | 109.6 | 0.120 | 11000 | 3.50e7 |

**Table 2** The same comparison as Table 1 in the presence of OP scattering

| +With OP | $I_{off}$ nA) | $C_{ins}$ (F/m) | $I_{ON}$ (uA) | $E_{FS}-E_I$ (eV) | $I_{ON}/I_{OFF}$ | $\langle v(0) \rangle$ cm/s |
|---|---|---|---|---|---|---|
| CNT | 10 | 8.42e-10 | 29.3 | 0.216 | 2900 | 2.72e7 |
| 1-layer GNR | 10 | 8.98e-10 | 24.4 | 0.260 | 2400 | 2.69e7 |
| 5-layer GNR | 10 | 1.27e-9 | 71.0 | 0.151 | 7100 | 3.30e7 |
| 10-layer GNR | 10 | 1.73e-9 | 105.4 | 0.118 | 10500 | 3.33e7 |

**Figure Captions**

Fig. 1. The schematic structure of the modeled multilayer GNRFET and the CNTFET. (a) The MOSFET has an intrinsic channel and heavily doped source and drain extensions. (b) The cross section of the wrapped-around gate GNRFET. (c) The cross section of the CNTFET with the coaxial gate.

Fig. 2. The atomistic structure of an AB-stacking multi-layer GNR. The z direction is defined along the GNR transport direction, x direction along the width direction, and y direction along the thickness direction. 'A' or 'B' in (b) denotes the atoms in A or B sublattice, respectively. And the subscript denotes the layer index in a multilayer GNR. The modeled multilayer GNR has armchair edges.

Fig. 3. (a) In the 'top of the barrier' ballistic transistor model, the $+k$ states are filled according to the source Fermi level $E_{FS}$, and the $-k$ states are filled according to the drain Fermi level $E_{FD}$. (b) In the presence of OP scattering, the $-k$ states are filled according to an effective Fermi level, $E_F^{'} = E_{FS} - \hbar\omega_{OP}$.

Fig. 4. (a) The bandstructures of 1-, 2-, 5- and 10-layer $n=22$ multi-layer AGNR with an interlayer coupling of 0.30 eV. The bandgap is 0.52 eV for 1-layer, 0.30 eV for 2-layer, 0.21 eV for 5-layer and 0.20 eV for 10-layer. (b) The bandgap as a function of interlayer coupling for 2-layer and 5-layer and 10-layer $n=22$ AGNRs.

Fig. 5. *Effect of the gate insulator*. The on current as a function of the gate oxide thickness (a) at the ballistic limit and (b) in the presence of OP scattering for GNRFETs and CNTFETs as shown in Fig. 1. The 1-, 5-, and 10-layer $n=22$ AGNR with a zero interlayer coupling have a similar bandgap as the simulated (20, 0) CNT with a $E_{g,CNT} \approx 0.50$ eV for a fair comparison. The gate insulator dielectric constant is $\kappa_{SiO2} \approx 3.9$. The smallest simulated value of the oxide thickness, $t_{ox} \approx 0.23$ nm is the equivalent value for a 3-nm thick $ZrO_2$ gate insulator with a dielectric constant of $\kappa \approx 25$, which results in the same gate insulator capacitance.

Fig. 6. *Effect of interlayer coupling*. The on-current as a function of the interlayer coupling for the 5-layer GNRFET (dashed lines) and the 2-layer GNRFET (solid lines) as shown in Fig. 1. A 3-nm-thick $ZrO_2$ gate insulator is used. The lines without symbols are computed for a ballistic channel and the lines with symbols are in the presence of OP scattering.

TOC fig. With a thin high-$\kappa$ wrapped-around gate and reduced interlayer coupling, the multilayer GNR FET can significantly outperform its CNT counterpart with a similar gate and bandgap in terms of the ballistic on-current

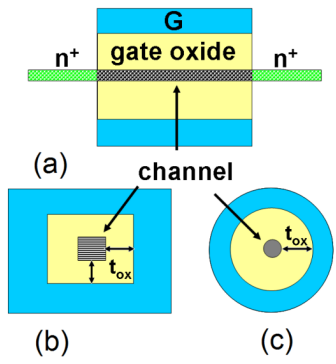

FIGURE 1

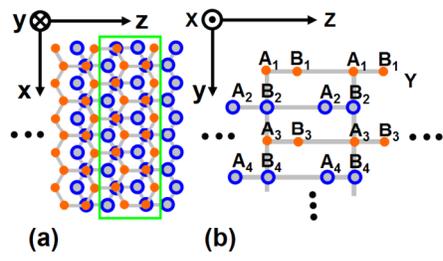

FIGURE 2

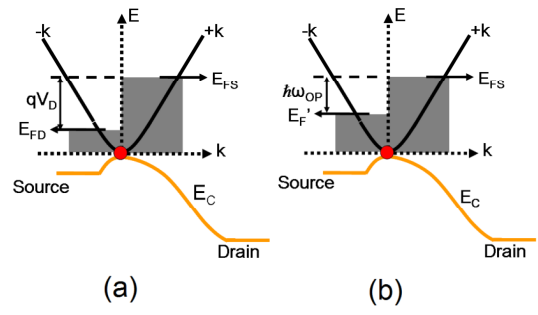

FIGURE 3

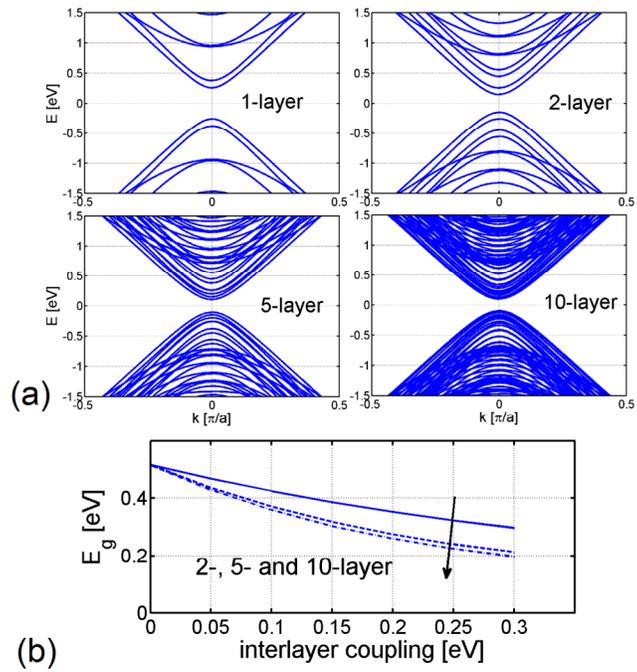

FIGURE 4

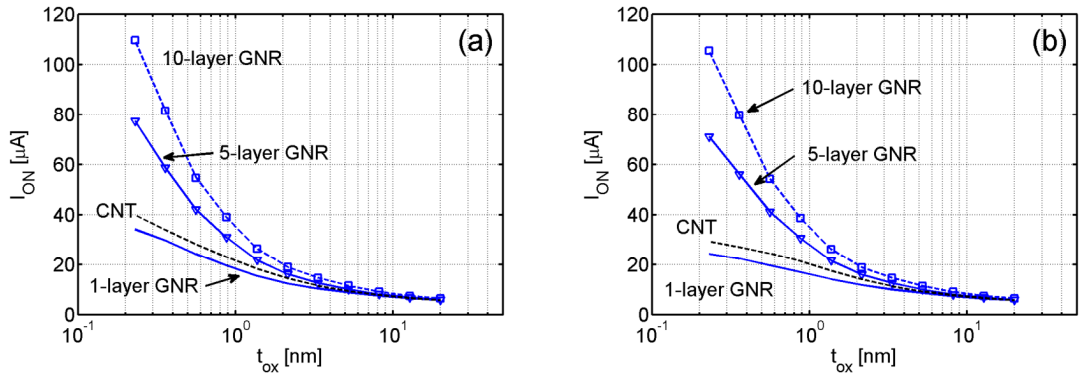

FIGURE 5

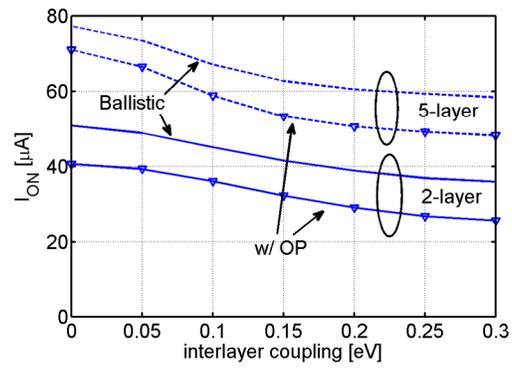

FIGURE 6